\documentclass[aps,pra,twocolumn,tightenlines,showpacs]{revtex4}
\usepackage{amssymb}
\usepackage{epsfig}
\usepackage{amsbsy}
\usepackage{amsmath}
\usepackage{graphicx}

\begin{document}

\title{Topological Phenomena in Trapped Ion Systems}
\author{T. Shi and J. I. Cirac}
\affiliation{Max-Planck-Institut f\"{u}r Quantenoptik, Hans-Kopfermann-Strasse 1, Garching, Germany}
\date{\today}

\begin{abstract}
We propose and analyze a scheme to observe topological phenomena with ions in microtraps. We consider a set of trapped ions forming a regular structure in two spatial dimensions and interacting with lasers. We find phonon bands with non-trivial topological properties, which are caused by the breaking of time reversal symmetry induced by the lasers. We investigate the appearance of edge modes, as well as their robustness against perturbations. Long-range hopping of phonons caused by the Coulomb interaction gives rise to flat bands which,  together with induced phonon-phonon interactions, can be used to produce and explore strongly correlated states. Furthermore, some of these ideas can also be implemented with cold atoms in optical lattices.
\end{abstract}
\maketitle


\section{Introduction}

Atomic systems have revealed themselves as ideal laboratories to study
many-body quantum systems. Atoms or molecules in optical lattices \cite%
{reviewBlochZwergerDalibard,bookLewenstein} realize Hubbard models, both
bosonic \cite{Jakschetal98} and fermionic \cite{HofstetterCiracZolleretc},
and thus can be used to study some of the most basic models of electron
transport and magnetism appearing in condensed matter physics. Starting from
the seminal experiment \cite{FirstMottInsulator}, many groups are now able
to implement different versions of Hubbard-like models. Although at the
moment the temperatures in most of those experiments are too high, single
site addressing \cite{BlochKuhr,Greiner} and other techniques are currently
opening up new avenues for the observation of many new and exciting
phenomena. Trapped ions can also realize bosonic Hubbard models \cite%
{PorrasCiracPRLHubbard}, as well as many other spin models \cite%
{PorrasCiracpaper,TIrev}. Despite recent experimental breakthroughs \cite%
{exp}, ions are not yet as versatile as atoms in lattices. However, this
situation may soon change with the development of surface traps \cite%
{Leibfried,Schmied,slusher,chuang,surfacetrap} or through the combination of
ions and optical lattices \cite{SchmiedCirac}.

One of the goals of atomic experiments is the simulation of the Fractional
Quantum Hall Effect (FQHE), one of the most celebrated discoveries in
condensed matter physics. In the standard electronic scenario, it appears at
very low temperatures in the presence of both an external magnetic field and
strong interactions. The first restricts the electron motion to the first
Landau level, whereas the latter is responsible for the generation of strong
correlations. It has recently been suggested that, under some conditions, in
lattice systems a similar behavior should occur even in the absence of a net
magnetic flux \cite{Haldane,flatbands}. The first condition is that the band
structure should possess a non-trivial topology, as characterized by the
so-called Chern Number (CN). The second one is that the width of the band
should be much smaller than the band gap, i.e., one has flat bands,
something which is associated to hopping beyond nearest neighboring sites.
In such a situation, the lattice problem resembles the standard scenario for
the appearance of the FQHE, whereby the flat band plays the role of the
first Landau level. Interaction energies larger than single particle ones
should thus give rise to strongly correlated states at low temperatures and,
in particular, to states displaying fractionalization as well as to anyonic
excitations.

Even in the absence of interactions, the non-trivial topology of the energy
bands gives rise to very intriguing behavior. In particular, when we
consider a problem with open boundary conditions (as opposed to periodic
boundary conditions) new energies appear within the band gap in the
spectrum, with a nearly linear dispersion relation. The corresponding modes
are located at the boundary of the system, and thus those are called
\textquotedblleft edge modes\textquotedblright\ \cite{SCZhangpaper}. The
presence of those modes is presumably very robust against perturbations \cite%
{Haldanebosonicpaper,robust}.

In atomic systems, in the bulk, one can simulate the action of a magnetic
field either by rotating the trap holding the atoms \cite{Cooper} or by
using laser fields \cite{Dum,Spielman,Dalibard}. In optical lattices, a very
promising alternative consists of using complex hopping amplitudes induced
by the lasers \cite{JakschandZoller} (see also \cite{Dalibard}). The
appearance of FQHE-like behavior in optical lattices was first considered by
Hafezi et al \cite{Hafezi} in the low density limit, where the lattice does
not play an important role. In addition, they proposed to use dipole-dipole
interactions to increase the band gap (see also \cite%
{robust,Cooper2005,Clark,Lewenstein}), and thus be able to increase the
density. In a recent paper, an implementation based on polar molecules \cite%
{GorshkovLukinZoller} is proposed which achieves very flat bands by
exploiting the dipole-dipole interaction among them, and thus is very well
suited for the observation of the FQHE.

In this paper we show that trapped ions can be used to investigate some of
the intriguing features mentioned above. We consider ions in microtraps,
forming a regular structure in two spatial dimensions. They are driven by
lasers, which are far off resonance with respect to an internal electronic
transition. By carefully choosing the laser configuration, the ions
experience a force that breaks the time-reversal symmetry. Furthermore,
phonons (i.e., motional quanta in each of the ions) can hop from one ion to
another due to the Coulomb interaction, which has a long range. As a result
of all that, we find a band structure with a non-trivial topological
structured, as witnessed by the corresponding CN. We investigate the
appearance of edge modes, as well as their fragility in the presence of
perturbations. We also study the regime of parameters where the bands are
very flat, whereby FQHE-like behavior could be investigated by adding
phonon-phonon interactions as proposed in \cite{PorrasCiracPRLHubbard}. Note
that other methods to obtain flat bands (albeit without topological
properties) have been recently proposed for trapped ions using
time-dependent gadgets \cite{Porrasnewwork}. We emphasize that the scheme
presented here is quite different from the one proposed for atoms in optical
lattices, as here the hopping terms are all real, whereas the on-site
Hamiltonian is the one that breaks the time-reversal symmetry. In fact, our
scheme resembles more that studied in the context of photonic edge modes
\cite{ReferenceFan}. The configuration we propose here could also be
suitable in setups involving atoms in optical lattices.

This paper is organized as follows: In Sec. II, we introduce the model setup
and derive an effective Hamiltonian for the phonons. In Sec. III, we analyze
the energy spectra of the system with toroidal, cylindrical, and planar
geometries. In particular, we derive the CN for the different bands and
relate them to the appearance of edge modes. In Sec. IV, we propose a scheme
to detect the edge modes by using a periodically driven electric field that
shakes the ion lattice. In Sec. V, we discuss that, due to the flat lowest
band, phonon-phonon interactions could lead to the observation of a bosonic
version of the FQHE. Some of the details of the derivations are presented in
the appendices.

\section{Model setup}

We consider $N$ ions confined in microtraps and interacting with lasers in a
planar geometry. The lasers are tuned far off-resonance with respect to any
optical transition, so that the ion stays most of the time in its ground
electronic internal state. Virtual transitions mediated by the lasers modify
the atomic motion. In particular, they give rise to a Doppler effect which,
when combined with dipole forces, generates something that resembles the
Lorenz force produced by a magnetic field. This force, which breaks the
time-reversal symmetry of the problem, is one of the main ingredients
required for the appearance of the topological features which we are looking
after. The other one is the Coulomb interaction among the ions, which allows
phonons to move from ion to ion \cite{PorrasCiracPRLHubbard}. The effective
dynamics of the phonons in the ion lattice are described by a Hamiltonian
which breaks time-reversal symmetry and, as we will see, has the desired
topological properties.

\begin{figure}[tbp]
\includegraphics[bb=8 212 586 783, width=8 cm, clip]{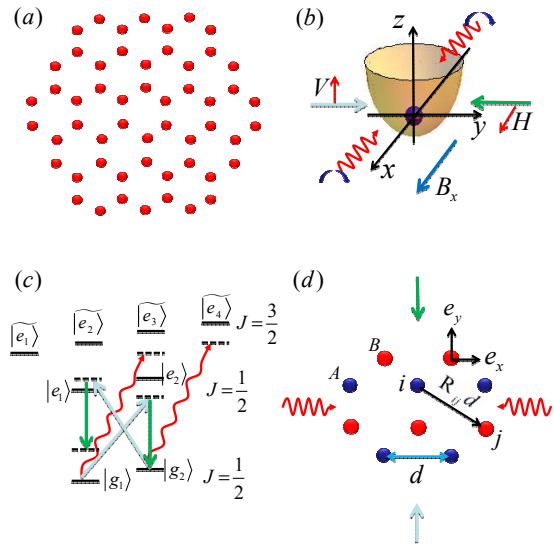}
\caption{(Color online) Configuration: (a) Ions are trapped in microtraps
forming a planar hexagonal geometry. (b) Laser and magnetic field
configuration for each ion in a microtrap. There is a homogenous magnetic
field as well as a laser standing-wave with circular polarization along the $%
x$ direction, and two counter-propagating lasers in lin$\perp$lin
configuration along the $y$ direction. (c) Ion internal structure, where the
quantization axis is chosen along the magnetic field ($x$ axis). For
simplicity, we consider no nuclear spin (hyperfine structure). The magnetic
field generates a Zeeman shift. The lasers propagating along the $x$
direction (in red) produce different AC-Stark shifts on the magnetic levels.
Lasers along the $y $ direction induce Raman transitions. (d) Detail of the
hexagonal lattice, with the unit vectors $e_{x,y}$ indicating the two
oscillation directions of the ions (vibrational modes). Different colors are
used for the $A$ and $B$ sublattices.}
\label{fig1}
\end{figure}

There are many lattice geometries that give rise to topological features.
Here, for the sake of concreteness, we will consider a hexagonal (honeycomb)
geometry [see Fig.\ref{fig1}(a)]. In this section we will first introduce
the interaction of the laser with each ion, and derive the corresponding
Lorenz-like force. Some of the details regarding a particular laser
configuration that implements those forces are given in Appendix A. Later
on, we will consider the interaction among the ions in order to derive the
effective Hamiltonian describing the phonon dynamics.

\subsection{Single-ion dynamics}

Let us first consider a single ion, as shown in Fig. \ref{fig1}b, with two
electronic (internal) ground levels $|g_{1,2}\rangle$ interacting with
lasers. The single-ion Hamiltonian contains three parts: $H_{\mathrm{si}}=H_{%
\mathrm{HO}}+H_{\mathrm{x}}^{\mathrm{las}}+H_{\mathrm{y}}^{\mathrm{las}}$
where ($\hbar=1$)
\begin{eqnarray}
H_{\mathrm{HO}} &=&\sum_{\alpha =x,y}(\frac{p_{\alpha }^{2}}{2M}+\frac{1}{2}%
M\omega _{\alpha }^{2}r_{\alpha }^{2}),  \notag \\
H_{\mathrm{x}}^{\mathrm{las}} &=&\Omega _{x}Kr_{x}\sigma _{z},  \notag \\
H_{\mathrm{y}}^{\mathrm{las}} &=&\frac{\Omega _{y}}{2}(e^{iKr_{y}}\sigma
^{+}+\mathrm{H.c.}),  \label{model}
\end{eqnarray}%
Here, $H_{\mathrm{HO}}$ describes the harmonic oscillation of the ion in the
plane, where $M$ is the mass, and $r_{\alpha }$, $p_{\alpha}$, and $\omega
_{\alpha }$ denote the position and momentum operators, and the trap
frequency along the direction $\alpha$, respectively. The terms $H_{\mathrm{%
x,y}}^{\mathrm{las}}$ describe internal level-dependent dipole and radiation
pressure forces, respectively. Here $K$ is twice the wavevector of the
lasers, $\Omega_{x,y}$ the effective Rabi frequencies, $\sigma
_{z}=\left\vert g_{2}\right\rangle \left\langle g_{2}\right\vert -\left\vert
g_{1}\right\rangle \left\langle g_{1}\right\vert $, and $\sigma
^{+}=\left\vert g_{2}\right\rangle \left\langle g_{1}\right\vert $.

Hamiltonian (\ref{model}) can be implemented in different ways. One possible
one is depicted in Fig. \ref{fig1}(b,c), which uses a magnetic field along
the $x$ direction, and two pairs of far off-resonant counterpropagating
beams along the $x$ and $y$ direction. In Appendix A we give the details of
this scheme, as well as how to obtain $H_{\mathrm{si}}$ after the adiabatic
elimination of the excited states.

It is convenient to transform the Hamiltonian (\ref{model}) according to the
unitary operator $U=\exp (iKr_{y}\sigma _{z}/2)$
\begin{equation}
\tilde{H}_{\mathrm{si}}=U^{\dagger }H_{\mathrm{si}}U=H_{\mathrm{HO}}+\frac{%
\Omega _{y}}{2}\sigma _{x}+(\Omega _{x}Kr_{x}+\frac{Kp_{y}}{2M})\sigma _{z},
\end{equation}%
where the last term represents the Doppler shift along the $y$ direction and
we have used the standard notation $\sigma _{x}=\sigma ^{+}+\sigma ^{-}$. In
the following we will simplify the model by assuming that $\left\vert \Omega
_{y}\pm \omega _{\alpha }\right\vert \gg \Omega _{x}\eta _{x}$, $\omega
_{y}\eta _{y}$, where $\eta _{\alpha }=K\ell _{\alpha }$ are the
dimensionless Lamb-Dicke parameters, which are typically $\leq 1$ (and $\ell
_{\alpha }=(2M\omega _{\alpha })^{-1/2})$ is the ground state size).
Assuming that the ion is initially prepared in the state $\left\vert
\downarrow \right\rangle =(\left\vert g_{2}\right\rangle -\left\vert
g_{1}\right\rangle )/\sqrt{2}$, we show in Appendix B that, after the
adiabatic elimination of the internal states, one ends up with the effective
Hamiltonian
\begin{equation}
H_{\mathrm{eff}}=\bar{H}_{\mathrm{HO}}+\Omega r_{x}p_{y}.  \label{Heff}
\end{equation}%
The position-momentum coupling is characterized by%
\begin{equation}
\Omega =-\sum_{\alpha =x,y}\frac{\Omega _{x}\Omega _{y}\omega _{\alpha }\eta
_{\alpha }^{2}}{\Omega _{y}^{2}-\omega _{\alpha }^{2}},
\end{equation}%
and $\bar{H}_{\mathrm{HO}}$ describes harmonic oscillations with
renormalized frequencies $\tilde{\omega}_{\alpha }$ (see Appendix B). This
Hamiltonian resembles the one describing the Lorenz force exerted by a
magnetic field on a charge (note, however, that it is not invariant under
rotations along the $z$ axis).

\subsection{Coulomb Interactions}

Let us now consider the whole set of ions. As mentioned above, the
microtraps are disposed according to a 2D honeycomb lattice, as shown in
Fig. \ref{fig1}(a,d). The interaction of each of the atoms with the lasers
is given by (\ref{Heff}). Coulomb forces induce repulsion between the ions.
Assuming that $d\gg \ell _{\alpha }$, we can expand the Coulomb interaction
up to second order in the displacements $r_{\alpha }^{(i)}$ about the center
of the i-th microtrap, obtaining
\begin{equation}
U_{\mathrm{C}}=\frac{e^{2}}{2d^{3}}\sum_{i,j,\alpha \beta }r_{\alpha
}^{(i)}U_{\alpha ,\beta }^{i,j}r_{\beta }^{(j)}
\end{equation}%
where
\begin{equation}
U_{\alpha ,\beta }^{i,j}=\left\{
\begin{array}{c}
\frac{1}{\left\vert \mathbf{R}_{ij}\right\vert ^{3}}[\mathbf{e}_{\alpha
}\cdot \mathbf{e}_{\beta }-\frac{3(\mathbf{e}_{\alpha }\cdot \mathbf{R}%
_{ij})(\mathbf{e}_{\beta }\cdot \mathbf{R}_{ij})}{\left\vert \mathbf{R}%
_{ij}\right\vert ^{2}}],\text{ \ \ \ \ \ \ \ \ \ }i\neq j, \\
\sum\limits_{j^{\prime }\neq i}\frac{1}{\left\vert \mathbf{R}_{ij^{\prime
}}\right\vert ^{3}}[\frac{3(\mathbf{e}_{\alpha }\cdot \mathbf{R}_{ij^{\prime
}})(\mathbf{e}_{\beta }\cdot \mathbf{R}_{ij^{\prime }})}{\left\vert \mathbf{R%
}_{ij^{\prime }}\right\vert ^{2}}-\mathbf{e}_{\alpha }\cdot \mathbf{e}%
_{\beta }],\text{ \ \ }i=j,%
\end{array}%
\right. .
\end{equation}

As shown in Fig. \ref{fig1}d, $\mathbf{e}_{\alpha }$ are the unit vectors,
and $\mathbf{R}_{ij}d$ is the vector pointing from the site $i$ to the site $%
j$. We have omitted the linear terms in $r_{\alpha }^{(i)}$, which just
redefine the trap center. Furthermore, the $U_{\alpha ,\alpha }^{i,i}$ just
shift the frequencies $\tilde{\omega}_{\alpha }$, and can be absorbed in $%
\tilde{\omega}_{\alpha }$.

The complete Hamiltonian, $H$, consisting of the single-ion terms as well as
the Coulomb interaction, can be rewritten in terms of the phonon
annihilation operators
\begin{equation}
a_{\alpha }^{(i)}=\sqrt{\frac{M\tilde{\omega}_{\alpha }}{2}}r_{\alpha
}^{(i)}+i\sqrt{\frac{1}{2M\tilde{\omega}_{\alpha }}}p_{\alpha }^{(i)}.
\label{a}
\end{equation}%
In the stiff limit $\beta _{\alpha }=e^{2}/(M\tilde{\omega}_{\alpha
}^{2}d^{3})\ll 1$, where we can drop the non-phonon number conserving terms
\cite{PorrasCiracpaper}, and we finally obtain the Hamiltonian for the
phonons
\begin{equation}
H=\sum_{ij,\alpha \beta }a_{\alpha }^{(i)\dagger }\mathcal{H}_{\alpha ,\beta
}^{i,j}a_{\beta }^{(j)}.  \label{H}
\end{equation}%
This describes phonon hopping without interaction (i.e., it is quadratic in
creation and annihilation operators), and is characterized by the
\textquotedblleft single-particle\textquotedblright\ Hamiltonian%
\begin{eqnarray}
\frac{\mathcal{H}_{\alpha ,\beta }^{i,j}}{\tilde{\omega}_{x}} &=&\frac{\beta
_{x}U_{\alpha ,\beta }^{i,j}}{2\gamma _{\alpha }\gamma _{\beta }}+\delta
_{ij}(\delta _{\alpha x}\delta _{\beta x}+\gamma _{y}^{2}\delta _{\alpha
y}\delta _{\beta y})  \notag \\
&&-iV_{\mathrm{b}}\delta _{ij}(\delta _{\alpha x}\delta _{\beta y}-\delta
_{\alpha y}\delta _{\beta x})  \label{2D}
\end{eqnarray}%
in terms of the dimensionless parameters $\beta _{x}$,$\ \gamma _{\alpha }=%
\sqrt{\tilde{\omega}_{\alpha }/\tilde{\omega}_{x}}$, and $V_{\mathrm{b}%
}=\Omega \gamma _{y}/(2\tilde{\omega}_{x})$. For the sake of simplicity, in
the following we will set $\gamma _{\alpha }=1$, so that the problem is
fully determined by two parameters. The corresponding values are restricted
by the conditions employed in the derivation of the Hamiltonian (\ref{2D}).
In particular, $\beta _{x}\ll 1$, and $\left\vert V_{b}\right\vert =\eta
^{2}\Omega _{x}/\Omega _{y}\alt\eta $, where we have taken $\eta _{x}=\eta
_{y}=\eta $. Taking the $^{20}$Ca$^{+}$ion, $\omega _{x}=\omega _{y}=\Omega
_{x}\sim 0.1MHz$, $\Omega _{y}\sim 0.3$-$1MHz$, and $d\sim 10$-$100\mu $m,
we obtain values for $\eta $ up to $0.8$ \cite{footnote}, $\beta _{x}\in
\lbrack 0.01,0.3]$ and $|V_{b}|\in \lbrack 0.05,0.3]$, which are the values
that we take in the examples given below.

The last term in Eq. (\ref{2D}) breaks the time-reversal symmetry, which
results from the position-momentum coupling. Thus, a non-trivial topological
phase may be expected to emerge in this setup.

\section{Topological properties}

In this section, we study the topological properties of the ion system. For
that, we first consider periodic boundary conditions (PBC), i.e., a toroidal
geometry, and analyze the spectrum of the Hamiltonian. Given the periodicity
of the problem, it has a band structure, where the existence of a
non-trivial topological structure of each band is characterized by the CN
\cite{SCZhangpaper}. We find that, depending on the values of $\beta _{x}$
and $V_{b}$, some of the bands exhibit non-zero CN. Then, we consider a
cylindrical geometry (i.e. PBC along one direction only). The spectrum is
very similar to the one corresponding to the torus, but now new energies
appear at the gaps between bands with certain CN configuration. The
corresponding eigenmodes (so-called edge modes \cite%
{SCZhangpaper,Volovikbook}) appear in pairs, are localized at the two
boundaries of the cylinder, and propagate in different directions. Finally,
we consider the experimentally relevant situation of full open boundary
conditions. In such a case, we also find edge modes. We further investigate
their robustness with respect to small random perturbations.

\subsection{Band structure and Chern numbers}

\begin{figure}[tbp]
\includegraphics[bb=14 433 585 775, width=8 cm, clip]{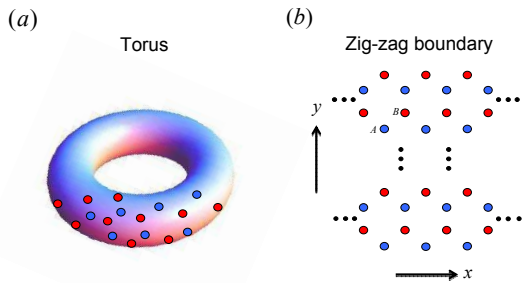}
\caption{(Color online) Different configurations considered in the text: (a)
torus (PBC); (b) cylinder.}
\label{fig3}
\end{figure}

In this subsection, we study the topological properties of the system with
PBC, namely, the system has translational invariance along both $x$ and $y$
directions [see Fig. \ref{fig3}(a)]. In order to find the single-particle
spectrum of $H$ (i.e. that of $\mathcal{H}$) we use the Fourier modes, and
thus diagonalize
\begin{equation}
\tilde{\mathcal{H}}_{\alpha ,\beta }({\mathbf{k}})=\frac{1}{N}\sum_{i,j}%
\mathcal{H}_{\alpha ,\beta }^{i,j}e^{-i\mathbf{k\cdot R}_{i,j}},
\end{equation}%
where $N$ is the number of sites and $\mathbf{k}=(k_{x},k_{y})$ is the
two-component quasimomentum in the first Brillouin zone (BZ).

\begin{figure}[tbp]
\includegraphics[bb=13 90 577 774, width=8 cm, clip]{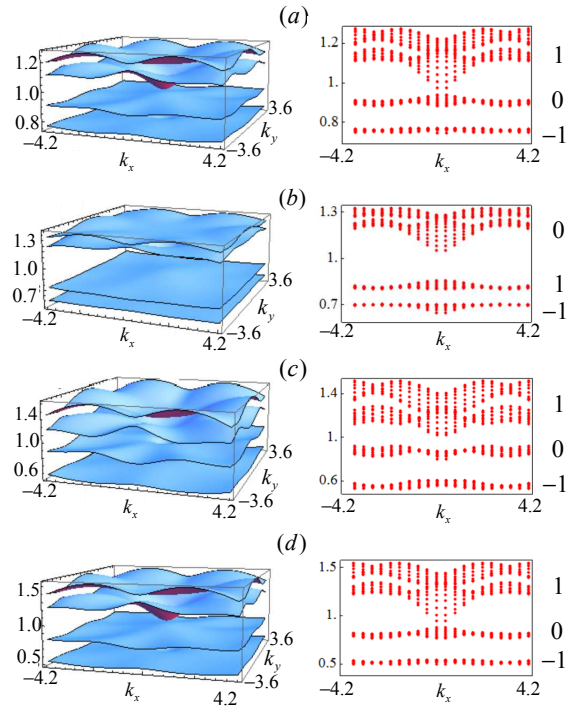}
\caption{(Color online) Single-particle energy spectra in units of $\tilde{%
\protect\omega}_x$ for the system with PBC and $N=400$. The left panels show
the energy spectra as a function of the quasimomenta $k_x$ and $k_y$ (in the
first Brillouin zone), whereas the right panels display the same spectra but
as a function of $k_x$ only, in order to highlight the different bands and
the corresponding gaps. The Chern numbers of each band are shown to the
right of such plots. (a) $\protect\beta_x=0.02$ and $V_{b}=-0.1$; (b) $%
\protect\beta_x=0.02$ and $V_{b}=-0.2$; (c) $\protect\beta_x=0.04$ and $%
V_{b}=-0.1 $; (d) $\protect\beta_x=0.04 $ and $V_{b}=-0.2$.}
\label{fig2}
\end{figure}

The energy spectrum of $\tilde{\mathcal{H}}_{\alpha ,\beta }({\mathbf{k}})$
is shown in Fig. \ref{fig2} for different parameters $\beta _{x}$ and $V_{%
\mathrm{b}}$. As one can see in the plot, there are a number of energy
bands. For $\beta _{x}=0.02$ and $V_{\mathrm{b}}=-0.1$, there is a big gap
between the first and the second bands, and a small one between the second
and the third bands. For $\beta _{x}=0.02$ and $V_{\mathrm{b}}=-0.2$, the
gap between the second and the third bands becomes larger, and the gap
between the first and the second one remains open. For larger $\beta
_{x}\sim 0.04$ and $V_{\mathrm{b}}=-0.1$, the band structure remains the
same. However, for $\beta _{x}\sim 0.04$ and $V_{\mathrm{b}}=-0.2$, the gap
between the second and the third bands becomes closed, whereas that between
the first and the second band remains.

The topological properties of each band, $s$, are characterized by the CN,
which is defined in terms of the eigenmodes of $\mathcal{H}$, $\left\vert s,%
\mathbf{k}\right\rangle $, in the limit $N\rightarrow \infty $ as \cite%
{SCZhangpaper}
\begin{equation}
C_{s}=\int_{\mathrm{BZ}}\frac{d^{2}k}{2\pi }\nabla _{\mathbf{k}}\times
\mathcal{A}_{\mathbf{k},s}.  \label{C}
\end{equation}%
Here, the integration is taken over quasimomenta in the first BZ, and $%
\mathcal{A}_{\mathbf{k}}=-i\left\langle s,\mathbf{k}\right\vert \partial _{%
\mathbf{k},s}\left\vert s,\mathbf{k}\right\rangle $. The CN can be
interpreted as the Berry phase obtained if we change adiabatically the
quasimomentum cyclically along the boundary of the BZ \cite{QN}.
Alternatively, if we interpret $\mathcal{A}_{\mathbf{k}}$ as a vector
potential, the $C$ is nothing but the total number of magnetic fluxes
through the whole BZ. Because the Brillouin zone forms a closed surface,
this number (and thus the CN) must be an integer, as it is required by the
single-valuedness of gauge transformations \cite{WuYangpaper}.

Given that the CN is an integer, we do not need to calculate it for $%
N\rightarrow \infty $, but a finite $N$ can already give us its value with
very high confidence. In general, however, it is very hard to determine the
CN through Eq. (\ref{C}) numerically, since $\mathcal{A}_{\mathbf{k}}$
contains singularities as long as $C\neq 0$. Thus, we use the alternative
expression \cite{TKNdN}
\begin{equation}
\tilde{C}_{s}=\text{Im}\int d^{2}k\sum_{s_{1}\leq s<s_{2}}\frac{(\partial
_{k_{x}}\mathcal{H}_{\mathbf{k}})_{s_{1},s_{2}}(\partial _{k_{y}}\mathcal{H}%
_{\mathbf{k}})_{s_{2},s_{1}}}{\pi (E_{\mathbf{k},s_{1}}-E_{\mathbf{k}%
,s_{2}})^{2}},  \label{C0}
\end{equation}%
where $s_{i}=1,2,\ldots $ denote the bands from the bottom to the top, $%
(\partial _{k_{i}}\mathcal{H}_{\mathbf{k}})_{s_{1},s_{2}}=\left\langle s_{1},%
\mathbf{k}\right\vert \partial _{k_{i}}\mathcal{H}_{\mathbf{k}}\left\vert
s_{2},\mathbf{k}\right\rangle $, and $E_{\mathbf{k},s_{i}}$ is the energy
corresponding to $\left\vert s_{i},\mathbf{k}\right\rangle $. Starting from $%
\tilde{C}_{s}$, we can recover the CN using%
\begin{equation}
\tilde{C}_{s}=\sum_{s^{\prime }=1}^{s}C_{s^{\prime }}.
\end{equation}

We have determined the CN of the three bands appearing in \ref{fig2}. For a
given $\beta _{x}$, as $\left\vert V_{\mathrm{b}}\right\vert $ increases,
the CN of the first, second and third bands vary from ($-1,0,1$) to ($-1,1,0$%
). A non-zero CN indicates the presence of a non--trivial topological
property in the system which, as we will see in the next subsection, is
reflected in the appearance of edge modes.

\subsection{Edge modes}

Based on the non-vanishing CN found in the previous section, we expect the
emergence of edge modes in the boundaries of our system. In this subsection,
we focus on those modes for two different geometries: cylinder [\ref{fig3}%
(b)](where there are PBC along the $x$-direction only), and plane [\ref{fig1}%
(a)].

\begin{figure}[tbp]
\includegraphics[bb=30 269 564 729, width=8 cm, clip]{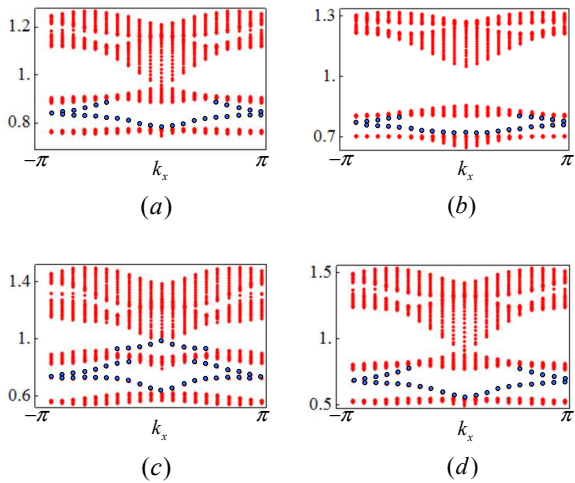}
\caption{(Color online) As in Fig.\protect\ref{fig2} for the cylinder
geometry and $N=400$: (a) $\protect\beta_x=0.02$ and $V_{b}=-0.1$; (b) $%
\protect\beta_x=0.02$ and $V_{b}=-0.2$; (c) $\protect\beta_x=0.04$ and $%
V_{b}=-0.1$; (d) $\protect\beta_x=0.04 $ and $V_{b}=-0.2$. The edge modes
are marked in blue.}
\label{fig4}
\end{figure}

For a cylindrical geometry, the translational invariance along the $y$
direction is broken. Still, given the periodicity along $x$, we can use
quasimomenta in this direction, and thus consider the single-particle
Hamiltonian
\begin{equation}
\tilde{\mathcal{H}}_{\alpha ,\beta }^{i_{y},j_{y}}(k_{x})=\frac{1}{N_{x}}%
\sum_{i_{x},j_{x}}\mathcal{H}_{\alpha ,\beta }^{i,j}e^{-ik_{x}(i_{x}-j_{x})},
\end{equation}%
where $i_{x,y}$ denote the $x$ and $y$ coordinates of ion $i$, $k_{x}=2\pi
n/N_{x}$ with $n\leq N_{x}$ integer, and $N_{x}$ are the number of sites in
a row. By numerical diagonalizing $\tilde{\mathcal{H}}_{\alpha ,\beta
}^{i_{y},j_{y}}(k_{x})$ for each value of $k_{x}$ we obtain the energy
spectra shown in Fig. \ref{fig4}. They are to be compared with those of Fig. %
\ref{fig2}. We choose two typical examples to explain the relation between
the CN and the edge modes. We first consider $\beta _{x}=0.02$ and $V_{%
\mathrm{b}}=-0.2$. The corresponding spectra are shown in Fig. \ref{fig2}(b)
and Fig. \ref{fig4}(b), respectively. As explained in the previous section,
the CN for the first, second, and third bands are $-1$, $1$, and $0$,
respectively. Thus, in the cylinder, two edge modes located in the up- and
bottom- boundaries emerge in the first band gap. They are signaled by the
blue dots that display a nearly linear dispersion relation with opposite
slopes, which indicate that they are counter-propagating. If we inspect the
band gap between the second and the third band, we see that no edge modes
show up. The reason is that total CN of the first and the second bands
(i.e., its sum) is zero. The second example is for $\beta _{x}=0.04$ and $V_{%
\mathrm{b}}=-0.1$ [Fig. \ref{fig4}(c)]. Here, the CN for the first, second,
and third bands are $-1$, $0$, and $1$, respectively. Therefore, in each
band gap there are two counter-propagating edge modes.

\begin{figure}[tbp]
\includegraphics[bb=21 527 579 734, width=8 cm, clip]{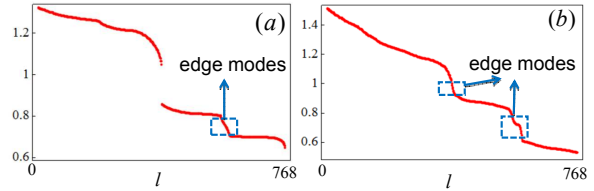}
\caption{(Color online) As in Fig.\protect\ref{fig2} for the planar geometry
and 384 sites: (a) $\protect\beta_x=0.02$ and $V_{b}=-0.2$; (b) $\protect%
\beta_x=0.04$ and $V_{b}=-0.1$. Here, $l$ denotes the eigenmode. Note that
for 384 sites, there are 768 eigenmodes. }
\label{fig5}
\end{figure}

\begin{figure}[tbp]
\includegraphics[bb=62 330 570 748, width=8 cm, clip]{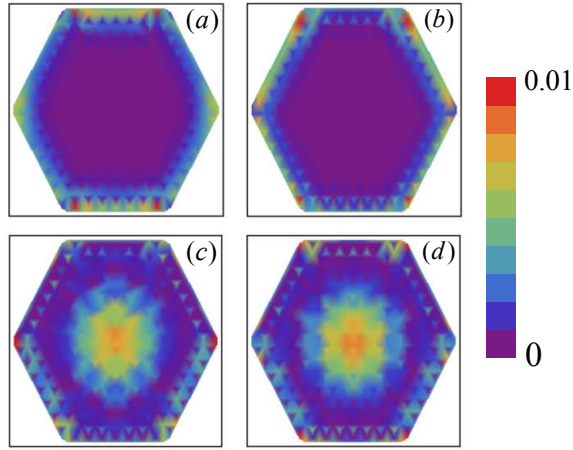}
\caption{(Color online) Spatial distribution for typical edge and bulk
eigenmodes corresponding to the system with planar geometry, with 384 sites,
$\protect\beta _{x}=0.04$ and $V_{b}=-0.1$. (a,c) For an energy $E=0.7\tilde{%
\protect\omega}_{x}$, corresponding to an edge mode and (c,d) $E= 0.9\tilde{%
\protect\omega}_{x}$, corresponding to a bulk mode; (a,c) $x$ component
(b,d) $y$ component.}
\label{fig6}
\end{figure}

For the planar geometry, the translational invariance is completely broken.
By numerical diagonalizing the Hamiltonian $\mathcal{H}$,
\begin{equation}
\sum_{j,\beta }\mathcal{H}_{\alpha ,\beta }^{i,j}u_{j,\beta
}^{l}=E_{l}u_{i,\alpha }^{l}.
\end{equation}%
we obtain the spectra of Fig. \ref{fig5} and the corresponding eigenmodes, $%
u^{l}$. As shown in Fig. \ref{fig5}(a), for $\beta _{x}=0.02$ and $V_{%
\mathrm{b}}=-0.2$, a set of edge modes emerges in the (small) first band
gap, while in the large second band gap there are none. Figure \ref{fig5}(b)
shows that edge modes emerge in both the first and second band gap for $%
\beta _{x}=0.04$ and $V_{\mathrm{b}}=-0.1$. Again, these results can be
understood in terms of the CN corresponding to the three bands. In order to
verify that those modes are located at the boundaries of our system, we plot
in Fig. \ref{fig6} the spatial distribution of a typical eigenmode
corresponding to the edge (a,b) and bulk (c,d). That is, for each ion $i$,
we plot at its location [i.e., at coordinates $(i_{x},i_{y})$] the value of $%
\left\vert u_{i,\alpha }^{l}\right\vert ^{2}$ for $\alpha =x$ (a,c) and $%
\alpha =y$ (b,d).

\begin{figure}[tbp]
\includegraphics[bb=17 238 580 732, width=8 cm, clip]{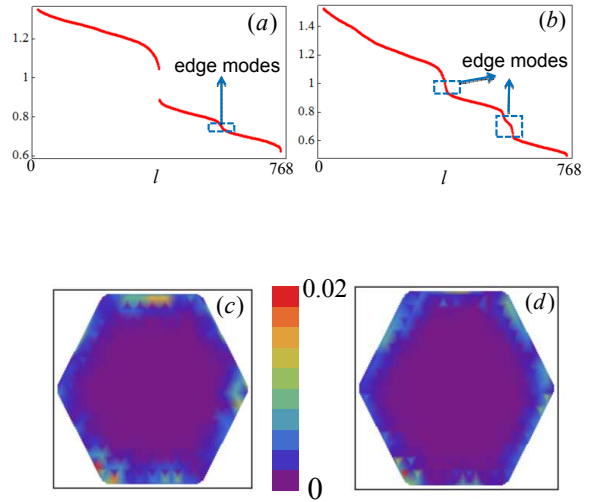}
\caption{(Color online) Single particle spectra in units of $\tilde{\protect%
\omega}_{x}$ (a,b) and the spatial distribution for the $x$ and $y$
components (c,d) of an eigenmode with energy $E=0.7\tilde{\protect\omega}%
_{x} $ for a planar geometry with 384 lattice sites. We have added to the
values $V_b$ and $\protect\beta_x$ random numbers as explained in the main
text.}
\label{fig7}
\end{figure}

Edge modes are supposed to be robust against small perturbations (with
energy densities smaller than the band gap; or, strictly speaking, as long
as the gap does not close) \cite{Haldanebosonicpaper}. One can understand
this behavior in a simple way by resorting to the cylindrical geometry.
There, impurities (or perturbations) cannot back-scatter the propagation of
phonons in one boundary since there are no modes available (they are located
in the other boundary of the cylinder). In order to test this robustness, we
have introduced random perturbations in the planar geometry. In particular,
we have added for each of the ions: (i) to $\left\vert V_{b}\right\vert $ a
random number in the interval $(0.15,0.25)$ and to $\tilde{\omega}_{x}$ a
random number in the interval $(0.95,1.05)$ in Fig. \ref{fig7}(a); (ii) to $%
\left\vert V_{b}\right\vert $ a random number in the interval $(0.05,0.15)$
and to $\tilde{\omega}_{x}$ a random number in the interval $(0.95,1.05)$ in
Fig. \ref{fig7}(b)-(d), which results in a random variation in $\beta _{x}$.
As we can see, in Fig. \ref{fig7}(a,b) the energy spectrum still has some
energies within the band gap, and the corresponding modes are located at the
boundary [Fig. \ref{fig7}(c,d)]. This confirms that the edge modes are
well-protected and robust.

\section{Probing the edge modes}

In this section, we propose a method to detect the edge modes. Since the
ions are charged, they can respond to an external electric field. We
consider a periodic electric field that shakes the lattice along the $x$%
-direction with frequency $\omega _{\mathrm{d}}$. If we initially prepare
the ions in the motional ground state (i.e., the state with no phonons),
phonons will be generated by the electric field. After a time $t_{\mathrm{f}%
} $, the field is turned off. As we will show below, if the frequency $%
\omega _{\mathrm{d}}$ is chosen to be the one corresponding to the edge
modes in the spectrum, phonons will be generated at the boundary of the
system, indicating the presence of edge modes at that frequency. The
presence of phonons can be detected using standard ion trap techniques \cite%
{polaronpaper,blackhole}.

The interaction between the ions and the electric field reads
\begin{equation}
H_{\mathrm{p}}(t)=\sum_{j}F_{j}r_{x}^{(j)}\cos (\omega _{\mathrm{d}}t),
\end{equation}%
where $F_{j}$ is the force driving the ion at site $j$. We define the
density distribution of the $x$-phonon at time $t$ as $\rho
_{j}(t)=\left\langle 0\right\vert a_{x}^{(j)\dagger
}(t)a_{x}^{(j)}(t)\left\vert 0\right\rangle $. Here, $\left\vert
0\right\rangle $ is the (vacuum) state with no phonons, and $a_{x}^{(j)}(t)$
denote the operators in the Heisenberg picture. By solving the Heisenberg
equations of motion, we obtain
\begin{equation}
\rho _{j}(t)=\frac{1}{4}\left\vert \sum_{l,m=\pm 1}u_{{j,x}}^{l}f_{l}^{\ast }%
\frac{e^{-iE_{l}t}-e^{im\omega _{\mathrm{d}}t}}{E_{l}+m\omega _{\mathrm{d}}}%
\right\vert ^{2},
\end{equation}%
where $f_{l}=\sum_{j}\Omega _{j}u_{{j,x}}^{l}$ is the effective driving
force corresponding to the eigenmode $l$, $\Omega _{j}=F_{j}/\sqrt{2M\tilde{%
\omega}_{x}}$.

\begin{figure}[tbp]
\includegraphics[bb=46 276 557 750, width=8 cm, clip]{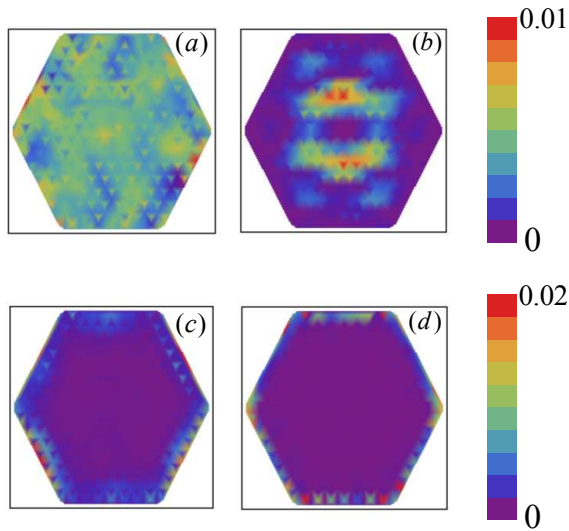}
\caption{(Color online) Density distribution of $x$-phonons after the
evolution time $t_{\mathrm{f}}$ for $\protect\beta _{x}=0.04$ and $%
V_{b}=-0.1 $: (a) $\protect\omega_{\mathrm{d}}=2.0\tilde{\protect\omega}_{x}$%
; (b) $\protect\omega_{\mathrm{d}}=0.6\tilde{\protect\omega}_{x}$; (c) $%
\protect\omega_{\mathrm{d}}=1.0\tilde{\protect\omega}_{x}$; (d) $\protect%
\omega_{\mathrm{d}}=0.7\tilde{\protect\omega}_{x}$.}
\label{fig8}
\end{figure}

In Fig. \ref{fig8}, we plot the normalized phonon density distribution
\begin{equation}
\bar{\rho}_j(t_{\mathrm{final}})=\frac{\rho_j(t_{\mathrm{final}})}{%
\sum_{j}\rho_j(t_{\mathrm{final}})}
\end{equation}%
for different driving frequencies $\omega _{\mathrm{d}}$ at time $t_{\mathrm{%
f}}=1000/\tilde{\omega}_{x}$. Here, we have considered the simplest case,
namely, $\Omega _{j}=\Omega _{0}$, the same for all ions. In Fig. \ref{fig8}%
(a), the driving frequency $\omega _{\mathrm{d}}=2.0\tilde{\omega}_{x}$ lies
outside the whole energy bands, and thus no resonance occurs, which results
in a weak phonon excitation. When the driving frequency $\omega _{\mathrm{d}%
}=0.6\tilde{\omega}_{x}$ is on resonance with one of the bulk states, many
phonons are generated in the bulk, as shown in Fig. \ref{fig8}(b). For
driving frequencies $\omega _{\mathrm{d}}=\tilde{\omega}_{x}$ and $0.7\tilde{%
\omega}_{x}$ some of the edge modes become resonant, and thus the density of
phonons in the boundary is much larger than that in the bulk, which
indicates the presence of an edge mode.

\section{phonon-phonon interaction}

As we showed in previous sections, with trapped ions it should be possible
to obtain bands in the spectrum with non trivial CN. Furthermore, due to the
nature of Coulomb interactions, phonons can hop over long distances, which
may give rise to flat bands. Thus, by engineering interactions among the
phonons one should be able to induce strong correlations and investigate
phenomena like fractionalization with this bosonic system. In this section
we will analyze the flatness of the bands for the model introduced in Sec.
II, as well as how to induce interactions among the phonons.

\begin{figure}[tbp]
\includegraphics[bb=23 454 569 761, width=8 cm, clip]{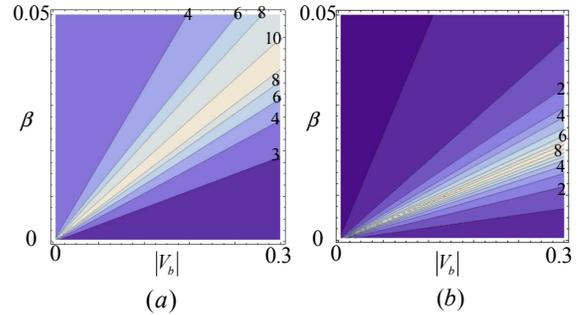}
\caption{(Color online) (a) Flatness $F$ of the first band as a function of $%
V_b$ and $\protect\beta_x$. (b) Same for a model where phonon hopping is
restricted to nearest neighbors.}
\label{fig9}
\end{figure}

The flatness, $F$, of a band can be characterized by the ratio between the
band gap and the bandwidth. In Fig. \ref{fig9}(a) we have plotted $F$ as a
function of $V_{b}$ and $\beta _{x}$ for the first band which has CN $-1$.
This figure shows that values of the order of 10 can be reached within our
setup. Those values should be large enough to allow for the investigation of
strong correlation effects with this system. In Fig. \ref{fig9}(b) we plot $%
F $ for the same model as the one considered here, but where phonon hopping
is (artificially) restricted to nearest neighbors. As one can see, the
flatness of the band becomes smaller than in the previous case, indicating
that the long range hopping present in the ion trap system helps to flatten
the band. Note that the model with NN hopping could be implemented using
atoms in optical lattices and, despite the fact that the flatness is
smaller, it may lead to interesting regimes for that system.

In Ref. \cite{PorrasCiracPRLHubbard} it was shown how a laser can induce
phonon-phonon interactions in the context of the Hubbard model. One can use
the same idea in our setup. In particular, in Appendix C we show how to
implement the interaction Hamiltonian
\begin{equation}
H_{\mathrm{U}}=U_{\mathrm{int}}\sum_{j}a_{x}^{(j)\dagger 2}a_{x}^{(j)2},
\end{equation}%
where $U_{\mathrm{int}}$ is determined by the Rabi frequency of the standing
wave and the Lamb-Dicke parameter. In order to enter the interesting regime
of strong correlations, the parameter $U_{\mathrm{int}}$ should be smaller
than the band gap, so that the interactions do not mix bands. However, it
should be larger than the bandwidth, so that the interactions dominate over
single phonon dynamics. Taking $\tilde{\omega}_{x}=0.1$MHz, $\beta _{x}=0.04$
and $V_{\mathrm{b}}=-0.2$, Fig. \ref{fig2}(d) shows that $U_{\mathrm{int}%
}\simeq 10$KHz should fulfill the conditions to obtain strong correlations.

\section{Conclusions and outlook}

In this paper, we have proposed a scheme to induce topological effects in a
system of ions trapped in microtraps. The main idea is to use laser fields
that break time-reversal symmetry. In particular, we have considered a pair
of counterpropagating lasers which are coupled to the position and momentum
of the ions, respectively, giving rise to a Lorenz-like force. This coupling
is very similar in spirit to the one proposed and used in neutral atom
systems. In a honeycomb lattice, one obtains a band structure with
non-trivial CN, which give rise to edge modes. We have analyzed the
robustness of such modes against small perturbations, an proposed a way for
detecting them. We have also studied the flatness of the bands, and show how
to induce phonon-phonon interactions. Our results indicate that as soon as
microtraps are available, they should be very well suited to investigate the
physics of the FQHE with them. We also emphasize that the setup used here to
induce the Lorenz-like force could be also used in the context of atoms in
optical lattices.

\acknowledgments

This project has been supported by the EU under the IP project AQUTE.

\appendix

\section{Physical Implementation}

In this appendix, we analyze a possible implementation of our model (\ref%
{model}). We consider a single earth alkali ion in a microtrap, in the
presence of a homogeneous magnetic field along the $x$ direction, as shown
in Fig. \ref{fig1}. For simplicity, we ignore the nuclear spin, so that the
ion has two electronic ground state levels, $\left\vert g_{n}\right\rangle $
($n=1,2$), and two sets of excited levels, $\left\vert e_{n}\right\rangle $ (%
$n=1,2$), and $\left\vert \tilde{e}_{n}\right\rangle $ ($n=1,\ldots ,4$),
corresponding to the fine splitting. We have chosen the quantization axis
along the direction of the magnetic field ($x$-direction); so, for instance,
$\left\vert g_{1,2}\right\rangle =\left\vert J=1/2,m_{J}=\mp
1/2\right\rangle _{x}$. We set the energy of the ground states as 0 and $%
\omega _{\mathrm{Z}}$ (Zeeman splitting), and that of the excited states as $%
\omega _{n}$, and $\tilde{\omega}_{n}$.

The ions interact with two sets of counterpropagating beams along the $x$
and $y$ directions. The first ones form a standing wave with frequency $%
\omega _{\mathrm{sw}}$ and circular polarization $\sigma ^{+}$, so that the
electric field is $\mathbf{E}_{0}\sin (K_{0}x+\phi )\cos (\omega _{\mathrm{sw%
}}t)$. They induce transitions $\left\vert g_{1}\right\rangle
\leftrightarrow \left\vert e_{2}\right\rangle $, $\left\vert \tilde{e}%
_{3}\right\rangle $, and $\left\vert g_{2}\right\rangle \leftrightarrow
\left\vert \tilde{e}_{4}\right\rangle $, which are described by the
Hamiltonian
\begin{eqnarray}
H_{\mathrm{sw}} &=&[\Omega _{\tilde{e}_{3}g_{1}}(x)\left\vert \tilde{e}%
_{3}\right\rangle \left\langle g_{1}\right\vert +\Omega
_{e_{2}g_{1}}(x)\left\vert e_{2}\right\rangle \left\langle g_{1}\right\vert
\notag \\
&&+\Omega _{\tilde{e}_{4}g_{2}}(x)\left\vert \tilde{e}_{4}\right\rangle
\left\langle g_{2}\right\vert +\mathrm{H.c.}]\cos (\omega _{\mathrm{sw}}t),
\end{eqnarray}%
where $\Omega _{ij}(x)=\Omega _{ij}\sin (K_{0}x+\phi )$, and the Rabi
frequencies $\Omega _{ij}=\left\langle i\right\vert \mathbf{d}\cdot \mathbf{E%
}_{0}\left\vert j\right\rangle $ are determined by the amplitude of the
electric field and the dipole moment $\mathbf{d}$ of ions, and $x$ is the
position operator of the atom in the $x$ direction.

The right-moving laser plane wave (propagating along the $y$ direction) is
vertically polarized (along $z$), has frequency $\omega _{\mathrm{R}}$, and
induces transitions $\left\vert g_{1}\right\rangle \leftrightarrow
\left\vert e_{2}\right\rangle $ and $\left\vert g_{2}\right\rangle
\leftrightarrow \left\vert e_{1}\right\rangle $, according to
\begin{equation}
H_{\mathrm{R}}=\frac{1}{2}(\Omega _{e_{2}g_{1}}^{\mathrm{R}}\left\vert
e_{2}\right\rangle \left\langle g_{1}\right\vert +\Omega _{e_{1}g_{2}}^{%
\mathrm{R}}\left\vert e_{1}\right\rangle \left\langle g_{2}\right\vert
)e^{iK_{0}y-i\omega _{\mathrm{R}}t}+\mathrm{H.c..}
\end{equation}%
where $\Omega _{ij}^{\mathrm{R}}$ are the Rabi frequencies, and $y$ is the
position operator of the atom in the $y$ direction. The left-moving plane
wave with frequency $\omega _{\mathrm{L}}$ and horizontal polarization
(along the $x$-direction) induces transitions $\left\vert g_{1}\right\rangle
\leftrightarrow \left\vert e_{1}\right\rangle $ and $\left\vert
g_{2}\right\rangle \leftrightarrow \left\vert e_{2}\right\rangle $,
according to%
\begin{equation}
H_{\mathrm{L}}=\frac{1}{2}(\Omega _{e_{1}g_{1}}^{\mathrm{L}}\left\vert
e_{1}\right\rangle \left\langle g_{1}\right\vert +\Omega _{e_{2}g_{2}}^{%
\mathrm{L}}\left\vert e_{2}\right\rangle \left\langle g_{2}\right\vert
)e^{-iK_{0}y-i\omega _{\mathrm{L}}t}+\mathrm{H.c..}
\end{equation}%
Note that we have used that the wavevectors of the standing and travelling
waves are approximately the same, $K_{0}$.

In the large detuning limit, where all the detunings are much larger than
the Rabi frequencies, by adiabatically eliminating the six higher energy
levels, we obtain the effective Hamiltonian%
\begin{eqnarray}
H_{\mathrm{inner}} &=&\sin ^{2}(K_{0}x+\phi )(\Delta _{1}\left\vert
g_{1}\right\rangle \left\langle g_{1}\right\vert -\Delta _{2}\left\vert
g_{2}\right\rangle \left\langle g_{2}\right\vert )  \notag \\
&&+\frac{\Omega _{y}}{2}\left\vert g_{2}\right\rangle \left\langle
g_{1}\right\vert e^{iKy}e^{-i\delta t}+\mathrm{H.c.},
\end{eqnarray}%
where $K=2K_{0}$, $\Omega _{y}=\Omega _{e_{2}g_{2}}^{\mathrm{L}\ast }\Omega
_{e_{2}g_{1}}^{\mathrm{R}}/[2(\omega _{\mathrm{R}}-\omega _{2})]$, and%
\begin{eqnarray}
\delta &=&\omega _{\mathrm{R}}-\omega _{Z}-\omega _{\mathrm{L}}+\frac{%
\left\vert \Omega _{e_{2}g_{1}}^{\mathrm{R}}\right\vert ^{2}-\left\vert
\Omega _{e_{2}g_{2}}^{\mathrm{L}}\right\vert ^{2}}{4(\omega _{\mathrm{R}%
}-\omega _{2})}  \notag \\
&\sim &\omega _{\mathrm{R}}-(\omega _{\mathrm{L}}+\omega _{\mathrm{Z}}),
\end{eqnarray}%
such that we can neglect all high-oscillating terms. The AC stark shifts%
\begin{eqnarray}
\Delta _{1} &=&\frac{1}{4}(\frac{\left\vert \Omega _{e_{2}g_{1}}\right\vert
^{2}}{\omega _{\mathrm{sw}}-\omega _{2}}-\frac{\left\vert \Omega _{\tilde{e}%
_{3}g_{1}}\right\vert ^{2}}{\tilde{\omega}_{3}-\omega _{\mathrm{sw}}}),
\notag \\
\Delta _{2} &=&\frac{\left\vert \Omega _{\tilde{e}_{4}g_{2}}\right\vert ^{2}%
}{4(\tilde{\omega}_{4}-\omega _{\mathrm{Z}}-\omega _{\mathrm{sw}})}
\end{eqnarray}%
can be taken to be the same, i.e., $\Delta _{1}=\Delta _{2}=\Delta $, if we
properly choose the frequency $\omega _{\mathrm{sw}}$. Taking the Lamb-Dicke
limit, where the ion displacements $r_{x}$ and $r_{y}$ along the $x$ and $y$
directions are much smaller than the optical wavelength, and assuming that
the trap center coincides with the node of the standing wave ($\phi =-\pi /4$%
) and $\delta =-\Delta $, the effective Hamiltonian becomes%
\begin{eqnarray}
H_{\mathrm{inner}} &=&\Omega _{x}Kr_{x}\sigma _{z}+\frac{\Omega _{y}}{2}%
(\sigma ^{+}e^{iKr_{y}}+\mathrm{H.c.})  \notag \\
&=&H_{\mathrm{x}}^{\mathrm{las}}+H_{\mathrm{y}}^{\mathrm{las}},
\end{eqnarray}%
where the effective Rabi frequency $\Omega _{x}=\Delta /2$. Notice that the
lattice spacing, $d$, fulfills $K_{0}d=4\pi n$, where $n$ is an integer.
This ensures that each ion experiences the same AC-Stark shift produced by
the lasers propagating in the $x$-direction (cf. $H_{\mathrm{x}}^{\mathrm{las%
}}$). For the second term, i.e., the Hamiltonian $H_{\mathrm{y}}^{\mathrm{las%
}}$ along $y$-direction, we replace $y$ by the displacement $r_{y}$, since
the phase $e^{iKy_{0}}$ generated by the coordinate $y_{0}$ of the
equilibrium position can be absorbed into the definition of $\sigma ^{+}$.

\section{Adiabatic elimination of the internal states}

In this Appendix, we show the details of the adiabatic elimination of the
internal states leading to Eq. (\ref{Heff}). Following the Fr\"{o}hlich
transformation \cite{frohlich}, we use the generator $S$ to rewrite the
Hamiltonian%
\begin{eqnarray}
H_{\mathrm{eff}} &=&e^{-S}\tilde{H}_{\mathrm{single}}e^{S}  \notag \\
&=&\tilde{H}_{\mathrm{single}}+[\tilde{H}_{\mathrm{single}},S]+\frac{1}{2}[[%
\tilde{H}_{\mathrm{single}},S],S]+...  \notag \\
&=&H_{0}+(H_{I}+[H_{0},S])+[H_{I}+\frac{1}{2}[H_{0},S],S]  \notag \\
&&+\frac{1}{2}[[H_{I},S],S],  \label{expan}
\end{eqnarray}%
where $H_{0}=H_{\mathrm{HO}}+\Omega _{y}\sigma _{z}/2$ and $H_{I}=-K(\Omega
_{x}r_{x}+p_{y}/2M)\sigma _{x}$. In order to make the notation more
transparent, we redefine the Pauli spin operators as follows: $\sigma
_{x}\rightarrow \sigma _{z}$ and $\sigma _{z}\rightarrow -\sigma _{x}$. By
requiring%
\begin{equation}
H_{I}+[H_{0},S]=0,
\end{equation}%
we obtain the generator%
\begin{eqnarray}
S &=&\Omega _{x}\eta _{x}(\frac{a_{x}\sigma _{+}}{\Omega _{y}-\omega _{x}}+%
\frac{a_{x}^{\dagger }\sigma _{+}}{\Omega _{y}+\omega _{x}})  \notag \\
&&-i\frac{1}{2}\omega _{y}\eta _{y}(\frac{a_{y}\sigma _{+}}{\Omega
_{y}-\omega _{y}}-\frac{a_{y}^{\dagger }\sigma _{+}}{\Omega _{y}+\omega _{y}}%
)-\mathrm{H.c..}
\end{eqnarray}%
Here, $a_{x,y}$ are defined according to Eq. (\ref{a}). In the limit $%
\left\vert \Omega _{y}\pm \omega _{\alpha }\right\vert \gg \Omega _{x}\eta
_{x}$, $\omega _{y}\eta _{y}$, we can keep up to second order in the
expansion (\ref{expan}) and obtain the effective Hamiltonian%
\begin{equation}
H_{\mathrm{eff}}=H_{0}+\frac{1}{2}[H_{I},S].
\end{equation}%
In that limit, if initially the internal state is the eigenstate $\left\vert
\downarrow \right\rangle _{z}$ of $\sigma _{z}$, we can project the
Hamiltonian $H_{\mathrm{eff}}$ onto that eigenstate and obtain%
\begin{equation}
H_{\mathrm{eff}}=\frac{p_{x}^{2}}{2M}+\frac{1}{2}M\tilde{\omega}%
_{x}^{2}r_{x}^{2}+\lambda _{y}(\frac{p_{y}^{2}}{2M}+\frac{1}{2}M\tilde{\omega%
}_{y}^{2}r_{y}^{2})+\Omega r_{x}p_{y},
\end{equation}%
where the renormalized frequencies $\tilde{\omega}_{x}=\lambda _{x}\omega
_{x}$ and $\tilde{\omega}_{y}=\omega _{y}/\sqrt{\lambda _{y}}$ are defined by%
\begin{eqnarray}
\lambda _{x} &=&\sqrt{1-\frac{2\Omega _{y}\Omega _{x}^{2}K^{2}}{M\omega
_{x}^{2}(\Omega _{y}^{2}-\omega _{x}^{2})}},  \notag \\
\lambda _{y} &=&1-\frac{\Omega _{y}K^{2}}{2M(\Omega _{y}^{2}-\omega _{y}^{2})%
},  \label{ly}
\end{eqnarray}%
and the position-momentum coupling is%
\begin{equation}
\Omega =-\sum_{\alpha =x,y}\frac{\Omega _{x}\Omega _{y}K^{2}}{2M(\Omega
_{y}^{2}-\omega _{\alpha }^{2})}.
\end{equation}%
To be consistent with the adiabatic limit, we can set $\lambda _{x}\sim
\lambda _{y}\sim 1$.

\section{Induction of phonon-phonon interactions}

In this Appendix, we show how one can induce phonon-phonon interactions by
using a laser. We basically follow the proposal of \cite%
{PorrasCiracPRLHubbard}.

An additional laser standing wave propagating along $x$-direction with
frequency $\tilde{\omega}_{\mathrm{sw}}$ and $\sigma ^{-}$ polarization
induces the transitions $\left\vert g_{1}\right\rangle \leftrightarrow
\left\vert \tilde{e}_{1}\right\rangle $ and $\left\vert g_{2}\right\rangle
\leftrightarrow \left\vert e_{1}\right\rangle $, $\left\vert \tilde{e}%
_{2}\right\rangle $. The light-ion interaction is%
\begin{eqnarray}
H_{\mathrm{int}} &=&[\tilde{\Omega}_{\tilde{e}_{1}g_{1}}(x)\left\vert \tilde{%
e}_{1}\right\rangle \left\langle g_{1}\right\vert +\tilde{\Omega}%
_{e_{1}g_{2}}(x)\left\vert e_{1}\right\rangle \left\langle g_{2}\right\vert
\notag \\
&&+\tilde{\Omega}_{\tilde{e}_{2}g_{2}}(x)\left\vert \tilde{e}%
_{2}\right\rangle \left\langle g_{2}\right\vert +\mathrm{H.c.})\cos (\tilde{%
\omega}_{\mathrm{sw}}t)\mathrm{,}
\end{eqnarray}%
where the Rabi frequencies
\begin{equation}
\tilde{\Omega}_{ij}(x)=\tilde{\Omega}_{ij}\cos (\tilde{K}_{0}x),
\end{equation}%
The adiabatic elimination of the electronic excited levels leads to the AC
stark shift%
\begin{equation}
H_{\mathrm{int}}=[\tilde{\Delta}_{1}\left\vert g_{1}\right\rangle
\left\langle g_{1}\right\vert +\tilde{\Delta}_{2}\left\vert
g_{2}\right\rangle \left\langle g_{2}\right\vert ]\cos ^{2}(\tilde{K}_{0}x),
\end{equation}%
where%
\begin{eqnarray}
\tilde{\Delta}_{1} &=&\frac{\left\vert \tilde{\Omega}_{\tilde{e}%
_{1}g_{1}}\right\vert ^{2}}{4(\tilde{\omega}_{\mathrm{sw}}-\tilde{\omega}%
_{1})}, \\
\tilde{\Delta}_{2} &=&\frac{1}{4}(\frac{\left\vert \tilde{\Omega}%
_{e_{1}g_{2}}\right\vert ^{2}}{\tilde{\omega}_{\mathrm{sw}}+\omega _{\mathrm{%
Z}}-\omega _{1}}+\frac{\left\vert \tilde{\Omega}_{\tilde{e}%
_{2}g_{2}}\right\vert ^{2}}{\tilde{\omega}_{\mathrm{sw}}+\omega _{\mathrm{Z}%
}-\tilde{\omega}_{2}}).  \notag
\end{eqnarray}%
Since $\tilde{\Delta}_{1}\tilde{\eta}^{2}$, $\tilde{\Delta}_{2}\tilde{\eta}%
^{2}\ll |\Omega _{y}|$ ($\tilde{\eta}=\tilde{K}_{0}\ell _{x}<1$ is the
Lamb-Dicke parameter), we can project $H_{\mathrm{int}}$ onto the ground
state $\left\vert \downarrow \right\rangle _{x}$ of $\sigma _{x}$, and obtain%
\begin{equation}
H_{\mathrm{int}}=\Omega _{\mathrm{int}}\cos ^{2}(\tilde{K}_{0}r_{x}),
\end{equation}%
where $\tilde{K}_{0}d=4\pi n$ ($n$ is an integer), $\Omega _{\mathrm{int}}$
is the effective Rabi frequency. By assuming that the center of the
microtrap is at the anti-node of the standing wave and expanding the cosine
term in $H_{\mathrm{int}}$, we obtain the phonon-phonon interaction%
\begin{equation}
H_{\mathrm{U}}=U_{\mathrm{int}}a_{x}^{\dagger 2}a_{x}^{2},
\end{equation}%
where we have neglected the phonon number non-conserving terms. Notice that
the second order in $r_{x}$ just renormalizes the trap frequency, and can be
absorbed in $\tilde{\omega}_{x}$. Here, the on-site interaction $U_{\mathrm{%
int}}=2\Omega _{\mathrm{int}}\tilde{\eta}^{4}$ can be tuned by the external
laser beams.


\begin{thebibliography}{99}
\bibitem{reviewBlochZwergerDalibard} I. Bloch, J. Dalibard, and W. Zwerger,
Rev. Mod. Phys. \textbf{80}, 885 (2008).

\bibitem{bookLewenstein} M. Lewenstein, A. Sanpera, and V. Ahufinger,
\textit{Ultracold Atoms in Optical Lattices: Simulating quantum many-body
systems }(Oxford University Press, 2012).

\bibitem{Jakschetal98} D. Jaksch, C. Bruder, J. I. Cirac, C. W. Gardiner,
and P. Zoller, Phys. Rev. Lett. \textbf{81}, 3108 (1998).

\bibitem{HofstetterCiracZolleretc} W. Hofstetter, J. I. Cirac, P. Zoller, E.
Demler, and M. D. Lukin, Phys. Rev. Lett. \textbf{89}, 220407 (2002).

\bibitem{FirstMottInsulator} M. Greiner, O. Mandel, T. Esslinger, T.W. H\"{a}%
nsch, and I. Bloch, Nature \textbf{415}, 39 (2002).

\bibitem{BlochKuhr} J. F. Sherson, C. Weitenberg, M. Endres, M. Cheneau, I.
Bloch, and S. Kuhr, Nature \textbf{467}, 68 (2010); C. Weitenberg, M.
Endres, J. F. Sherson, M. Cheneau, P. Schau\ss , T. Fukuhara, I. Bloch, and\
S. Kuhr, Nature \textbf{471}, 319 (2011).

\bibitem{Greiner} W. S. Bakr, A. Peng, M. E. Tai, R. Ma, J. Simon, J.
Gillen, S. Foelling, L. Pollet, and M. Greiner, Science \textbf{329}, 547
(2010); W. S. Bakr, J. I. Gillen, A. Peng, S. Foelling, and M. Greiner,
Nature \textbf{462}, 74 (2009).

\bibitem{PorrasCiracPRLHubbard} D. Porras and J. I. Cirac, Phys. Rev. Lett.
\textbf{93}, 263602 (2004).

\bibitem{PorrasCiracpaper} D. Porras and J. I. Cirac, Phys. Rev. Lett.
\textbf{92}, 207901 (2004); X. L. Deng, D. Porras, and J. I. Cirac, Phys.
Rev. A \textbf{72}, 063407 (2005). K. Kim, M. S. Chang, R. Islam, S.
Korenblit, L. M. Duan, and C. Monroe, Phys. Rev. Lett. \textbf{103}, 120502
(2009).

\bibitem{TIrev} see R. Blatt and C. F. Roos, Nat. Phys. \textbf{8}, 277
(2012) and references therein.

\bibitem{exp} A. Friedenauer, H. Schmitz, J. T. Glueckert, D. Porras and T.
Sch\"{a}tz, Nat. Phys. \textbf{4}, 757 (2008); Ch. Schneider, M. Enderlein,
T. Huber and T. Sch\"{a}tz, Nat. Photon. \textbf{4}, 772 (2010); K. Kim, M.
S. Chang, S. Korenblit, R. Islam, E. E. Edwards, J. K. Freericks, G. D. Lin,
L. M. Duan, and C. Monroe, Nature \textbf{465}, 590 (2010); R. Islam, E. E.
Edwards, K. Kim, S. Korenblit, C. Noh, H. Carmichael, G. D. Lin, L. M. Duan,
C. C. J. Wang, J. K. Freericks, and C. Monroe, Nat. Commun. \textbf{2}, 377
(2011); B. P. Lanyon, C. Hempel, D. Nigg, M. M\"{u}ller, R. Gerritsma, F. Z%
\"{a}hringer, P. Schindler, J. T. Barreiro, M. Rambach, G. Kirchmair, M.
Hennrich, P. Zoller, R. Blatt, and C. F. Roos, Science \textbf{7}, 57 (2011).

\bibitem{Leibfried} J. Chiaverini, R. B. Blakestad, J. Britton, J. D. Jost,
C. Langer, D. Leibfried, R. Ozeri, and D. J. Wineland, Quant. Inf. Comput.
\textbf{5}, 419 (2005).

\bibitem{Schmied} R. Schmied, J. H. Wesenberg, and D. Leibfried, Phys. Rev.
Lett. \textbf{102}, 233002 (2009).

\bibitem{slusher} J. T. Merrill, C. Volin, D. Landgren, J. M. Amini, K.
Wright, S. C. Doret, C. S. Pai, H. Hayden, T. Killian, D. Faircloth, K. R.
Brown, A. W. Harter, and R. E. Slusher, New J. Phys. \textbf{13}, 103005
(2011).

\bibitem{chuang} S. X. Wang, J. Labaziewicz, Y. Ge, R. Shewmon, and I. L.
Chuang, Phys. Rev. A \textbf{81}, 062332 (2010); T. H. Kim, P. F. Herskind,
T. Kim, J. Kim, and I. L. Chuang, Phys. Rev. A \textbf{82}, 043412 (2010);
R. J. Clark, Z. Lin, K. S. Diab, and I. L. Chuang, arXiv:1009.0036; T. H.
Kim, P. F. Herskind, and I. L. Chuang, Appl. Phys. Lett. \textbf{98}, 214103
(2011).

\bibitem{surfacetrap} A. Ozakin and F. Shaikh, arXiv:1109.2160; D. L.
Moehring, C. Highstrete, D. Stick, K. M. Fortier, R. Haltli, C. Tigges, and
M. G. Blain, New J. Phys. \textbf{13}, 075018 (2011).

\bibitem{SchmiedCirac} R. Schmied, T. Roscilde, V. Murg, D. Porras, and J.
I. Cirac, New J. Phys. \textbf{10}, 045017 (2008).

\bibitem{Haldane} F. D. M. Haldane, Phys. Rev. Lett. \textbf{61}, 2015
(1988).

\bibitem{flatbands} K. Sun, Z. Gu, H. Katsura, and S. Das Sarma, Phys. Rev.
Lett. \textbf{106}, 236803. (2011); E. Tang, J. W. Mei, and X. G. Wen, Phys.
Rev. Lett. \textbf{106}, 236802 (2011); T. Neupert, L. Santos, C. Chamon,
and C. Mudry, Phys. Rev. Lett. \textbf{106}, 236804 (2011); Y. Wang, H. Yao,
Z. Gu, C. D. Gong, and D. N. Sheng, Phys. Rev. Lett. \textbf{108}, 126805
(2012). S. Yang, Z. C. Gu, K. Sun, and S. Das Sarma, arXiv:1205.5792.

\bibitem{SCZhangpaper} X. L. Qi, T. L. Hughes, and S. C. Zhang, Phys. Rev. B
\textbf{78}, 195424 (2008).

\bibitem{Haldanebosonicpaper} F. D. M. Haldane and S. Raghu, Phys. Rev.
Lett. \textbf{100}, 013904 (2008); Phys. Rev. A \textbf{78}, 033834 (2008).

\bibitem{robust} N. Y. Yao, C. R. Laumann, A. V. Gorshkov, H. Weimer, L.
Jiang, J. I. Cirac, P. Zoller, and M. D. Lukin, arXiv:1110.3788.

\bibitem{Cooper} N. R. Cooper, N. K. Wilkin and J. M. F. Gunn, Phys. Rev.
Lett. \textbf{87}, 120405 (2001).

\bibitem{Dum} R. Dum and M. Olshanii, Phys. Rev. Lett. \textbf{76}, 1788
(1996).

\bibitem{Spielman} Y. J. Lin, R. L. Compton, A. R. Perry, W. D. Phillips, J.
V. Porto, and I. B. Spielman, Phys. Rev. Lett. \textbf{102}, 130401 (2009).

\bibitem{Dalibard} J. Dalibard, F. Gerbier, G. Juzeli\={u}nas, and P. \"{O}%
hberg, Rev. Mod. Phys. \textbf{83}, 1523 (2011).

\bibitem{JakschandZoller} D. Jaksch and P. Zoller, New J. Phys. \textbf{5},
56 (2003).

\bibitem{Hafezi} M. Hafezi, A. S. S\o rensen, E. Demler, and M. D. Lukin,
Phys. Rev. A \textbf{76}, 023613 (2007).

\bibitem{Cooper2005} N. R. Cooper, E. H. Rezayi and S. H. Simon, Phys. Rev.
Lett. \textbf{95}, 200402 (2005).

\bibitem{Clark} S. G. Bhongale, L. Mathey, S. W. Tsai, C. W. Clark, and E.
Zhao, Phys. Rev. Lett. \textbf{108}, 145301 (2012).

\bibitem{Lewenstein} B. C. Sansone, C. Trefzger, M. Lewenstein, P. Zoller,
and G. Pupillo, Phys. Rev. Lett. \textbf{104}, 125301 (2010); C. Menotti, C.
Trefzger, and M. Lewenstein, Phys. Rev. Lett. \textbf{98}, 235301 (2007).

\bibitem{GorshkovLukinZoller} N. Y. Yao, C. R. Laumann, A. V. Gorshkov, S.
D. Bennett, E. Demler, P. Zoller, and M. D. Lukin, arXiv:1207.4479.

\bibitem{Porrasnewwork} A. Bermudez, T. Schaetz, and D. Porras, Phys. Rev.
Lett. \textbf{107}, 150501, (2011).

\bibitem{ReferenceFan} K. Fang, Z. Yu, and S. Fan, Phys. Rev. B \textbf{84},
075477 (2011).

\bibitem{footnote} Since we are using a Raman transition (see Appendix A),
the Lamb-Dicke parameter is a factor 2 larger than the standard one.

\bibitem{Volovikbook} G.E. Volovik, \textit{The Universe in a Helium Droplet}
(Clarendon, Oxford, 2003).

\bibitem{QN} D. Xiao, M. Chang, and Q. Niu, Rev. Mod. Phys. \textbf{82},
1959 (2010).

\bibitem{WuYangpaper} T. T. Wu and C. N. Yang, Phys. Rev. D \textbf{12},
3845 (1975).

\bibitem{TKNdN} D. J. Thouless, M. Kohmoto, M. P. Nightingale, and M. den
Nijs, Phys. Rev. Lett. \textbf{49}, 405 (1982).

\bibitem{polaronpaper} V. M. Stojanovic, T. Shi, C. Bruder, and J. I. Cirac,
arXiv:1206.7010.

\bibitem{blackhole} B. Horstmann, B. Reznik, S. Fagnocchi, and J. I. Cirac,
Phys. Rev. Lett. \textbf{104}, 250403 (2010); B. Horstmann, R. Sch\"{u}%
tzhold, B. Reznik, S. Fagnocchi, and J. I. Cirac, New J. Phys. \textbf{13},
045008 (2011).

\bibitem{frohlich} H. Fr\"{o}hlich, Proc. Roy. Soc. A \textbf{215}, 291
(1952).
\end{thebibliography}
\end{document}